\documentclass[usegraphicx,usenatbib,natbib,useAMS]{mn2e}

\usepackage{times}     %use Adobe fonts - nicer for viewing PDF
\usepackage{aas_macros}% for journal abbreviations in Bibtex references from ADS
\usepackage{amssymb}   % for math symbols

\def\halpha{\mbox{H$\alpha$}}
\def\hbeta{\mbox{H$\beta$}}
\def\hgamma{\mbox{H$\gamma$}}
\def\hdelta{\mbox{H$\delta$}}

\def\ecs{\mbox{\,erg~s$^{-1}$~cm$^{-2}$}}

\def\ecsa{\mbox{\,erg~cm$^{-2}$~s$^{-1}$~{\AA}$^{-1}$}}
\def\lesssim{\mathrel{\hbox{\rlap{\hbox{\lower4pt\hbox{$\sim$}}}\hbox{$<$}}}}
\def\gtrsim{\mathrel{\hbox{\rlap{\hbox{\lower4pt\hbox{$\sim$}}}\hbox{$>$}}}}

\DeclareRobustCommand{\ion}[2]{%
\relax\ifmmode
 \ifx\testbx\f@series
  {\mathbf{#1\,\mathsc{#2}}}\else
  {\mathrm{#1\,\mathsc{#2}}}\fi
 \else\textup{#1\,{\mdseries\textsc{#2}}}%
\fi}

\title[X-shooter Observations of CSWA\,5]{X-shooter Observations of
  the Gravitational Lens System CASSOWARY 5 
\thanks{Based on public data from the X-shooter commissioning
  observations collected at the European Southern Observatory
  VLT/Melipal telescope, Paranal, Chile.}}

\author[L. Christensen et al.]
       {Lise Christensen\thanks{lise.christensen@ph.tum.de}$^{1,2}$ ,
         Sandro D'Odorico$^2$, Max Pettini$^{3,4}$, Vasily Belokurov$^3$,
         \newauthor N. Wyn Evans$^3$, Melinda Kellogg$^5$ and Jo\"{e}l Vernet$^2$\\
        $^1$ Excellence Cluster Universe, Technische Universit\"{a}t
        M\"{u}nchen, Bolzmanstrasse 2, 85748 Garching, Germany\\
        $^2$ European Southern Observatory, Karl-Schwarzschild-Strasse 2, 85748 Garching bei M\"{u}nchen, Germany\\
        $^3$ Institute of Astronomy, Madingley Rd, Cambridge, CB3 0HA, UK\\
        $^4$ International Centre for Radio Astronomy Research,
        University of Western Australia, 35 Stirling Hwy, Crawley,
        WA 6009, Australia\\
        $^5$ The University of Virginia's College at Wise, 1 College Avenue, Wise, VA 24293, USA\\
\\ 
}
\date{Accepted 2010 April 15. Received 2010 March 24; in original form 2010 January 20}
      
\pubyear{2010}

\begin{document}

\maketitle

\label{firstpage}

\begin{abstract}
We confirm an eighth gravitational lens system in the CAmbridge Sloan
Survey Of Wide ARcs in the skY (CASSOWARY) catalogue.  Exploratory
observations with the X-shooter spectrograph on the Very Large
Telescope (VLT) of the European Southern Observatory (ESO) show the
system CSWA\,5 to consist of at least three images of a blue
star-forming galaxy at $z = 1.0686$, lensed by an apparent foreground
group of red galaxies one of which is at $z = 0.3877$.  The lensed
galaxy exhibits a rich spectrum with broad interstellar absorption
lines and a wealth of nebular emission lines. Preliminary analysis of
these features shows the galaxy to be young, with an age of $\sim
25$--50\,Myr. With a star-formation rate of $\sim
20$\,M$_\odot$~yr$^{-1}$, the galaxy has already assembled a stellar
mass $M_\ast \sim 3 \times 10^9$\,M$_\odot$ and reached half-solar
metallicity.  Its blue spectral energy distribution and Balmer line
ratios suggest negligible internal dust extinction.  A more in-depth
analysis of the properties of this system is currently hampered by the
lack of a viable lensing model. However, it is already clear that
CSWA\,5 shares many of its physical characteristics with the general
population of UV-selected galaxies at redshifts $z = 1$--3, motivating
further study of both the source and the foreground mass concentration
responsible for the gravitational lensing.
\end{abstract}


%%%%%%%%%%%%%%%%%%%%%%%%%%%%%%%%%%%%%%%%%%%%%%%%%%%%%%%%%%%%%%%%%%%%%%
%%%%%%%%%%%%%%%%%%%%%%%%%%%%%%%%%%%%%%%%%%%%%%%%%%%%%%%%%%%%%%%%%%%%%%
\section{Introduction}

\begin{figure*}
\begin{center}
\vspace{-0.5cm}
\includegraphics[width=12.5cm]{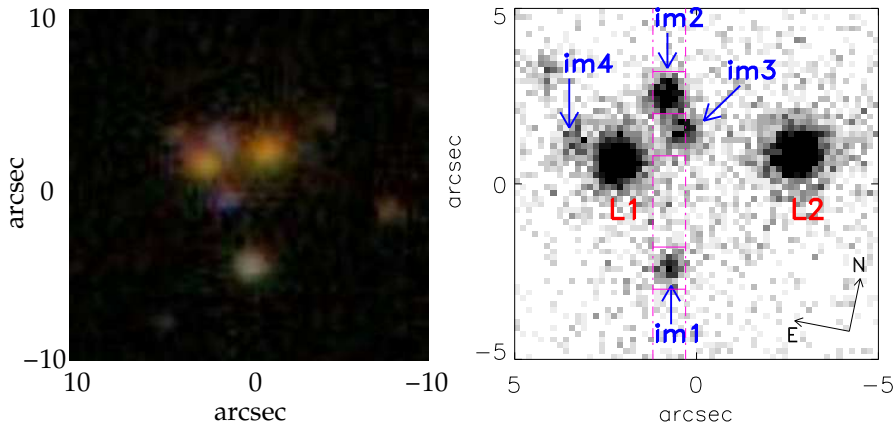}
\end{center}
\vspace{-0.25cm}
\caption{\textit{Left: } $20\times20$\,arcsec SDSS $g$, $r$, $i$
  colour composite image of the CSWA\,5 system.~ \textit{Right: }
  $r^\prime$-band image of CSWA\,5 recorded with a 20\,s long exposure
  of the X-shooter acquisition and guiding camera, at position angle
  P.A.\,=\,8\,degrees.  The scale of the right hand image is extended
  by a factor of two relative to the left one.  In the right-hand
  figure, the lensed images of the source are labelled im1--im4,
  reflecting the sequence of light arrival times (on the assumption
  that im4 is one of the lensed images).  Two foreground massive red
  galaxies are labelled L1 and L2. Note that the blue image im4 is not
  clearly visible in the SDSS composite on the left.  The X-shooter
  slit is overlaid on the right-hand image of the field; we have also
  indicated the apertures used in the extraction of the
  one-dimensional spectra of im1, im2, and im3, as described in
  Section~\ref{sect:data}.  Coordinates of all six labelled objects
  are given in Table~\ref{tab:maggies}.  }
\label{fig:SDSS_im}
\end{figure*}

\begin{table*}
 \begin{minipage}{150mm}
    \caption{SDSS coordinates and magnitudes for the different components
    of the CSWA\,5 lens system.}
%  \centering
  \begin{tabular}{lllccccc}
\hline
Object & RA (J2000) & Dec (J2000) & $~~~u~~~$ & $g$ & $r$ & $i$ & $z$ \\
\hline
\hline
L1  & ~~12:44:51.35~~ & ~~01:06:43.0~~ & ~~22.34$\pm$0.50~~ & ~~20.61$\pm$0.04~~ & 
      ~~19.12$\pm$0.02~~ & ~~18.44$\pm$0.02~~ & ~~18.02$\pm$0.04~~ \\
L2  & ~~12:44:51.00~~ & ~~01:06:44.0~~ & 22.33$\pm$0.79 & 20.43$\pm$0.06 & 
      18.44$\pm$0.02 & 17.87$\pm$0.01 & 17.67$\pm$0.04 \\
im1 & ~~12:44:51.22~~ & ~~01:06:40.1~~ & 22.07$\pm$0.20 & 21.74$\pm$0.06 & 
      22.48$\pm$0.17 & 23.85$\pm$0.67 & 21.75$\pm$0.45 \\
im2 & ~~12:44:51.27~~ & ~~01:06:45.2~~ & 20.79$\pm$0.08 & 21.09$\pm$0.04 & 
      20.91$\pm$0.05 & 20.67$\pm$0.07 & 20.42$\pm$0.19 \\
im3 & ~~12:44:51.23~~ & ~~01:06:44.2~~ & \\
im4 & ~~12:44:51.44~~ & ~~01:06:43.5~~ &  \\
\hline
\end{tabular}
\label{tab:maggies}

\textit{Notes.}~ The magnitudes quoted are SDSS model magnitudes
    and are considerably more uncertain than the formal errors 
    quoted because of blending.
    No SDSS magnitudes are available for images im3
    and im4.\\
    ~~~~The coordinates of the objects were determined from the
    X-shooter acquisition image (see Figure~\ref{fig:SDSS_im}),
    relative to the SDSS photometric source
    catalogue. Positional uncertainties are $\sim 0.1$\,arcsec.

\end{minipage}
\end{table*}

Galaxies at redshifts $z \gtrsim 1$ are generally too faint for their
rest-frame ultraviolet (UV) spectra to be recorded with the high
resolution and signal-to-noise (S/N) ratios required for an in-depth
study of their most important physical properties.  Consequently, our
knowledge of the high redshift galaxy population is still largely
based on analyses of stacked spectra which average together
observations of many galaxies in order to improve the S/N ratio
\citep[e.g.][]{Shapley03, Vanzella09, weiner09}.  A few exceptions are
sources which, thanks to their fortuitous location behind foreground
mass concentrations such as massive galaxies, or groups and clusters
of galaxies, are strongly lensed so that their fluxes are
significantly boosted.  Observations of what has become the archetypal
strongly lensed galaxy, MS\,1512-cB58 (cB58 for short) at $z=2.7276$
\citep{pettini00,teplitz00,pettini02,baker04,siana08}, have provided
clear demonstrations of the power of this technique in giving insights
into the interstellar medium and stellar populations of distant
galaxies with a level of detail which would otherwise be unobtainable
until the next generation of 30+\,m optical-infrared telescopes.

The large area of sky sampled by the Sloan Digital Sky Survey
\citep[SDSS,][]{york00} has allowed systematic searches for strongly
lensed galaxies \citep[e.g.][]{allam07,lin09}, with the dual interest
of identifying and studying the most massive galaxies in the lensing
objects, and distant galaxies in the lensed sources. The Sloan Lens
ACS Survey, or SLACS,\footnote{\tt http://www.slacs.org/} targets
galaxies with emission lines at more than one redshift, or redshifted
nebular emission lines superimposed on the spectra of early type
galaxies.  These candidate gravitationally lensed systems are then
followed up with high spatial resolution images from the Advanced
Camera for Surveys on the \textit{Hubble Space Telescope}
\citep[\textit{HST;} e.g.][]{bolton06,willis05}.  In a complementary
approach, The CAmbridge Sloan Survey Of Wide ARcs in the skY
(CASSOWARY) targets multiple, blue companions around massive
ellipticals in the SDSS photometric catalogue as likely candidates for
wide-separation gravitational lens systems \citep[see][]{belokurov09}.

These and other search strategies have led to numerous recent
discoveries of strongly lensed galaxies at $z \gtrsim 1$ which have
been followed up in detail over a range of wavelengths \citep[see][and
  references
  therein]{lemoine03,cabanac05,belokurov07,coppin07,swinbank07,cabanac08,stark08,finkelstein09,hainline09,
  quider09,siana09,dessauges09,quider10}.  However, all of these
studies have of necessity focused only on limited wavelength
intervals, mostly in the rest-frame UV (redshifted to optical
wavelengths) or the rest frame optical (redshifted into the
near-infrared).  It is only with the recent advent of the X-shooter
spectrograph on the Very Large Telescope (VLT) facility of the
European Southern Observatory (ESO), that both wavelength ranges can
now be recorded at once with good sensitivity. The obvious advantages
have been demonstrated in our recent study of CASSOWARY\,20
\citep[CSWA\,20,][]{pettini10}, a blue star-forming galaxy at $z_{\rm
  em} = 1.433$ lensed into a wide separation ($\sim 6$\,arcsec)
Einstein Cross by a massive luminous red galaxy at $z_{\rm abs} =
0.741$ with velocity dispersion $\sigma_{\rm lens} \simeq
500$\,km~s$^{-1}$.

\begin{figure*}
\begin{center}
\includegraphics[width=17cm, bb=100 375 1020 700, clip]{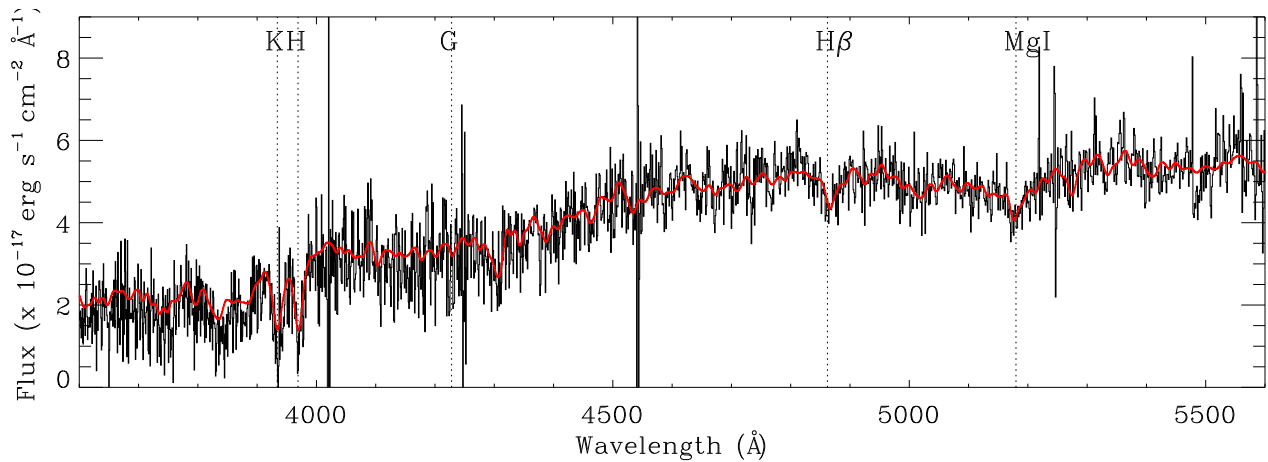}
\end{center}
\caption{SDSS spectrum of galaxy L2 at $z_{\mathrm{abs}}=0.3877$
  (shifted to the rest frame). The best fit stellar template with a
  velocity dispersion of $\sigma = 278$\,km~s$^{-1}$ is shown in
  red. }
\label{fig:SDSS_spec}
\end{figure*}

The current CASSOWARY catalogue\footnote {\tt
  http://www.ast.cam.ac.uk/research/cassowary/} includes seven
spectroscopically confirmed lensed galaxies at $z > 1$.  In this paper
we confirm an eighth case with X-shooter observations of CSWA\,5.  As
can be seen from Figure~\ref{fig:SDSS_im}, this multiple system
consists of two foreground red galaxies one of which (L2) has an SDSS
spectrum which shows it to be at $z_{\rm abs} = 0.3877$, and four
fainter sources, separated by $\sim 1$--5\,arcsec.  Three of these
images, im1, im2 and im3, are shown by the observations reported here
to be gravitationally lensed images of an emission line galaxy at
$z_{\rm em} = 1.0686$, while the fourth image, not covered by our
observations, remains to be confirmed spectroscopically.  Coordinates
and magnitudes of the different components of CSWA\,5 are given in
Table~\ref{tab:maggies}.  However, we caution that the SDSS pipeline
magnitudes, which are based on exponential profile fits, are likely to
be significantly more uncertain than formal errors quoted, because of
blending between the different components of this complex system.  In
particular, while image im1 may appear to be bluer than im2 from the
magnitudes listed in Table~\ref{tab:maggies}, we suspect that this is
just due to blending with the foreground red galaxy L1, because our
X-shooter spectra of all three images, im1, im2, and im3, exhibit the
same continuum slope.

This paper is organised as follows.
We describe the X-shooter observations in Section~\ref{sect:data} and
consider the lensing galaxies in Section~\ref{sect:lens}.  The main
focus of the paper is the rich absorption and emission line spectrum
of the source, described in Sections~\ref{sect:source} and
\ref{sect:phys_con} respectively.  We discuss our main results in
Section~\ref{sec:discuss} and conclude in Section~\ref{sec:conclude}.
Unless otherwise stated, we assume a flat cosmological model with
$H_0=73$~km~s$^{-1}$~Mpc$^{-1}$ and $\Omega_{\Lambda}$~=~0.72
\citep{spergel07}.

%%%%%%%%%%%%%%%%%%%%%%%%%%%%%%%%%%%%%%%%%%%%%%%%%%%%%%%%%%%%%%%%%%%%%%
%%%%%%%%%%%%%%%%%%%%%%%%%%%%%%%%%%%%%%%%%%%%%%%%%%%%%%%%%%%%%%%%%%%%%%
\section{Observations and Data Reduction}
\label{sect:data}

The first of the second generation VLT instruments, X-shooter
\citep{dodorico06} is a three arm, single object echelle spectrograph
which was successfully integrated into standard VLT operation on
October 1, 2009.  The instrument covers simultaneously the wavelength
range from 300\,nm to 2400\,nm at resolving powers $R =
\lambda/\Delta\lambda = 5100$, 8800 and 5600 in the UV-B ($\Delta
\lambda = 300$--550\,nm), VIS-R ($\Delta \lambda = 550$--1015\,nm),
and NIR arms ($\Delta \lambda = 1025$--2400\,nm) respectively.  The
corresponding slit widths are 1.0, 0.9, and 0.9\,arcsec respectively.
Although X-shooter is optimised for observations of single targets
with a 11\,arcsec long slit, it also allows for simultaneous recording
of close pairs of objects on the sky.  As a test of the performance of
the spectrograph, two CASSOWARY candidates were observed during the
commissioning run of the instrument in March 2009.  Results on the
first object, CSWA\,20, have recently been published by
\citet{pettini10}.  This second paper reports the observations of
CSWA\,5.

Referring to Figure~\ref{fig:SDSS_im}, the entrance slit of X-shooter
was rotated to a position angle on the sky of P.A.\,=\,8\,degrees and
aligned so as to capture the light of images im1 and im2; at this PA,
part of the light from image im3 also falls into the slit.  The UV-B
and VIS-R detectors were binned by a factor of two in the spectral
direction.  The total exposure time on source was 2400\,s, split into
two 1200\,s long integrations; each integration on source was followed
by a 1200\,s exposure on a nearby blank sky region.  From the
$r^\prime$-band acquisition frame reproduced in
Figure~\ref{fig:SDSS_im}, we measured a seeing full width at half
maximum FWHM\,=\, 0\farcs9.  The seeing improved during the course of
the observations; from the spatial extent of the spectra themselves we
measured FWHM\,=\,0\farcs8 in the UV-B arm, 0\farcs7 in the VIS-R arm,
and 0\farcs6 in the NIR arm.  With the chosen setup, the slit losses
in the three arms are estimated to be 15--10\%.

The spectra were reduced with a preliminary version of the ESO
X-shooter pipeline \citep{goldoni06}, which uses subtraction of the
sky lines based on the procedures developed by \citet{kelson03}.  For
the UV-B and VIS-R data, we found it most advantageous to subtract the
sky signal from the same two-dimensional (2D) frames on which the
light from CSWA\,5 was recorded, while for the NIR data sky
subtraction from the separate blank sky 2D frames worked best.  The
pipeline reduction used calibration spectra taken during the
commissioning run for order location and tracing, flat fielding and
wavelength calibration. The final product from the pipeline is an
extracted 2D, wavelength calibrated, rectified spectrum with orders
combined using a weighting scheme. For further data processing and
analysis we used standard {\sc iraf} tools. We produced 1D spectra
using a 1.2\,arcsec extraction aperture in all three arms and for all
three images of the source (see right panel of
Figure~\ref{fig:SDSS_im}).  As discussed in Section~\ref{sect:source},
the 16 strongest emission lines which were clearly detected in all
three spectra were measured to be at the same redshift within $\Delta
z = 5 \times 10^{-5}$.  Furthermore, we found no significant
differences in the emission line ratios nor in the continuum slopes of
the spectra of images im1, im2 and im3, leading us to conclude that
all three images are lensed counterparts of the same background
galaxy. Thus, unless otherwise specified, the following analysis was
performed on the sum of the spectra of im1, im2 and im3, so as to
maximise the S/N ratio of the data.

Absolute flux calibration was performed with reference to the
\textit{HST} white dwarf standard GD\,71, while the O8\,V star
Hipparcos 69892 provided a reference smooth spectrum for dividing out
telluric absorption. Both stars were observed the night prior to the
observations of CSWA\,5. Even though Hipparcos 69892 was observed at
similar airmass as CSWA\,5, the correction for atmospheric absorption
is not perfect, especially in the regions of low transmission between
the $J$, $H$, and $K$ bands.  A correction for atmospheric extinction
was applied assuming a standard atmospheric extinction curve.
Overall, we expect the absolute flux calibration of these
commissioning data to be accurate to within $\sim 20\%$, based on the
comparison between the spectrum and the SDSS magnitudes of im1 (the
least contaminated of the four images).

%%%%%%%%%%%%%%%%%%%%%%%%%%%%%%%%%%%%%%%%%%%%%%%%%%%%%%%%%%%%%%%%%%%%%%
%%%%%%%%%%%%%%%%%%%%%%%%%%%%%%%%%%%%%%%%%%%%%%%%%%%%%%%%%%%%%%%%%%%%%%
\section{Foreground Galaxies}
\label{sect:lens}

Our X-shooter observations described above did not include either of
the (presumably) foreground red galaxies visible in
Figure~\ref{fig:SDSS_im}. Galaxy L2 has an SDSS spectrum reproduced in
Figure~\ref{fig:SDSS_spec}.  We identify a number of stellar
absorption features, labelled in Figure~\ref{fig:SDSS_spec}, which
indicate that it lies at redshift $z_{\rm abs} = 0.3877$.  In order to
obtain an estimate of the velocity dispersion (and hence the mass) of
the galaxy, we used the the penalised pixel fitting method of
\citet{cappellari04}, taking into account the difference in velocity
resolution between the input stellar template spectra of
\citet{sanchez-blazquez06} and the SDSS spectrum.  The best fit gives
a stellar velocity dispersion $\sigma_{\mathrm{L2}} = 278 \pm
6$\,km~s$^{-1}$, where the uncertainty reflects the dispersion in the
values of $\sigma$ obtained with different stellar templates.  The
corresponding mass is $M(r) = \pi \sigma^2 r/G$ (appropriate to an
isothermal sphere), or $M_{\rm L2} \simeq 6 \times 10^{11} M_{\odot}$,
adopting $r=10$\,kpc .

\begin{figure}
{\hspace*{-0.9cm}\includegraphics[width=9.25cm]{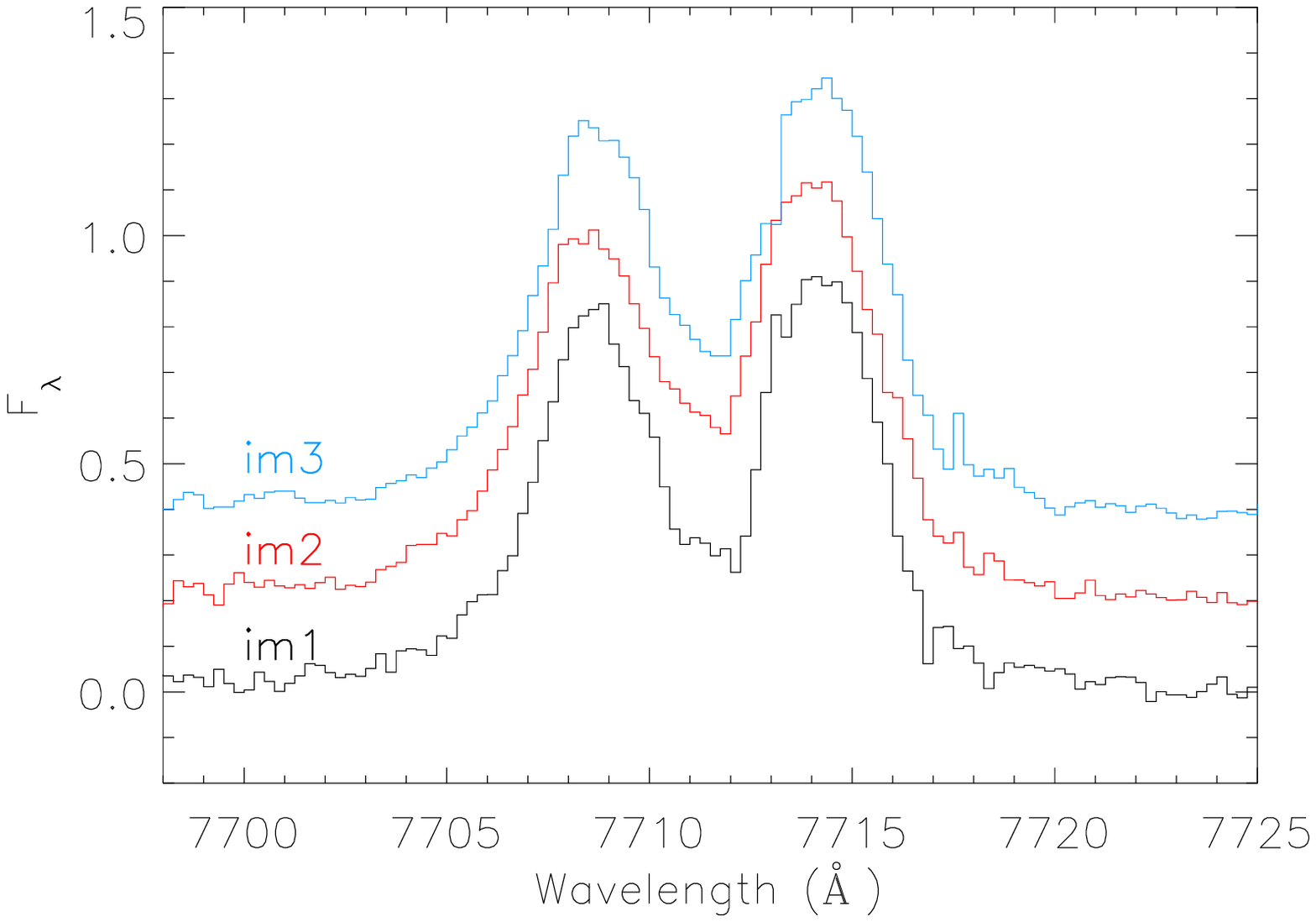}}
{\hspace*{-0.9cm}\includegraphics[width=9.25cm]{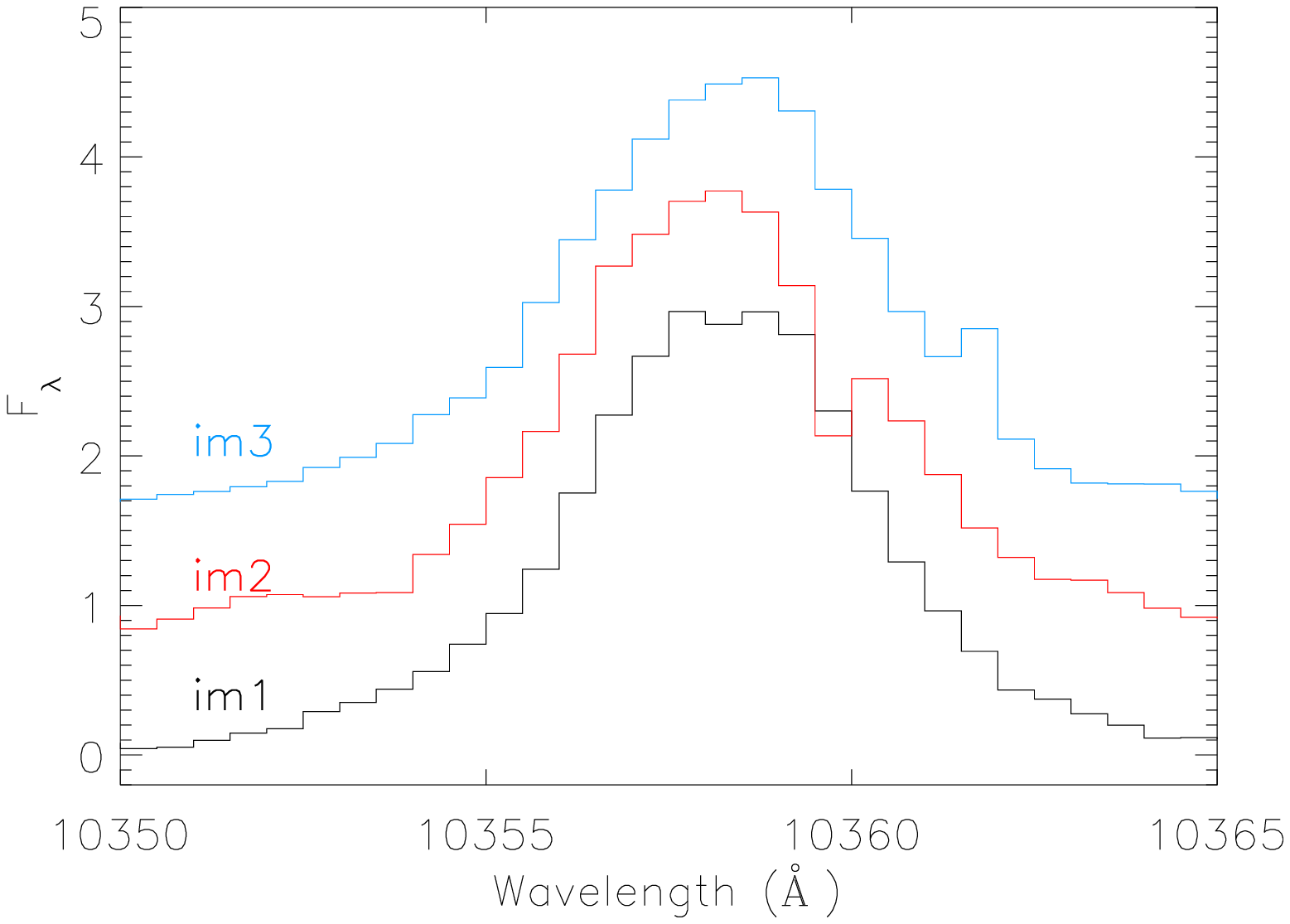}}
\caption{Profiles of the 
[\ion{O}{ii]}\,$\lambda \lambda 3728,3730$ (upper panel) 
and [\ion{O}{iii]}\,$\lambda 5008$ (lower panel) emission lines
extracted separately from images im1, im2, and im 3.
The spectra
have been offset vertically, and scaled to facilitate comparison.
The line profiles are indistinguishable within the accuracy
of the data, indicating that the three images
are lensed counterparts of the same background galaxy at 
$z_{\rm em} = 1.06856$.  }
\label{fig:3OIIlines}
\end{figure}

At redshift $z=0.3877$, the SDSS $z$ filter is closest to rest-frame
$r$.  From the SDSS measured magnitude $z({\rm L2}) = 17.67$, we
deduce an absolute magnitude $M_r({\rm L2}) = -23.17$, given the
distance modulus $m - M = 41.17$ in our cosmology and applying an
evolutionary correction of $0.85 z$ to $M_r$, as proposed by
\citet{bernardi06}.  There is no SDSS spectrum for galaxy L1. If it is
at the same redshift as L2, $z=0.3877$, its absolute magnitude is
$M_r({\rm L1}) = -22.82$ and the projected separation between the two
galaxies is 28\,kpc. The absolute magnitude and velocity dispersion of
L2 are within the range of values found by \citet{bernardi06} in their
sample of massive early-type galaxies selected from the SDSS database.

\begin{figure*}
\begin{center}
{\hspace{0.15cm}\includegraphics[width=15.005cm]{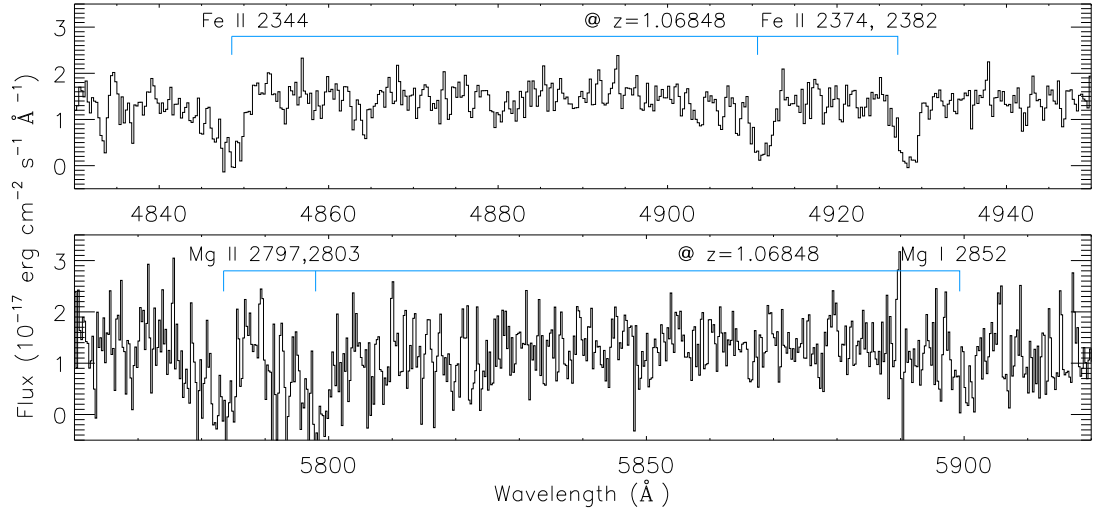}}
\end{center}
\caption{Examples of interstellar absorption lines in the X-shooter
  spectrum of CSWA\,5. The full list of lines identified is given in
  Table~\ref{tab:abslines}.  }
\label{fig:abslines}
\end{figure*}

\subsection{Lens Model}
\label{sec:lens_model}

For the interpretation of many of the properties of the source, it
would be advantageous to obtain estimates of the magnification factors
which apply to images im1--im3.  To this end, we attempted a simple
modelling of the CSWA\,5 system following the approach described in
\citet{pettini10}, but quickly realised that we have insufficient data
at our disposal for even the simplest characterization of the lens.
For the idealised case of an isolated singular isothermal sphere, the
typical deflection is given by:
\begin{equation}
\Delta \theta = 1\farcs15 \left( {\sigma
\over 200 {\rm \,km\,s}^{-1}}\right)^2 
\left( {D_{\rm ds} \over D_{\rm s}} \right)
\label{eq:isodef}
\end{equation}
where $D_{\rm ds}$ is the angular diameter distance between deflector
and source, whilst $D_{\rm s}$ is the distance between observer and
source.  If galaxy L1 is: (a) the lens, and (b) at the same redshift
as L2, then applying the measured $\sigma_{\mathrm{L2}} = 278
$\,km~s$^{-1}$ (which would actually be an upper limit for L1, since
L1 is fainter than L2---see Table~\ref{tab:maggies}) to
eq.~(\ref{eq:isodef}) would result in typical deflections of $\sim
1.4$\,arcsec, or only about half the deflections of images im1--im4.
We are thus led to one of two conclusions.  Either L1 is substantially
more massive than L2, as would be case if it is at a higher redshift,
or L2 and other galaxies in the group significantly boost the lensing
by L1.  Presumably the second option is the more likely, given the
proximity of L1 and L2 on the sky and their broadly similar magnitudes
and colours (Table~\ref{tab:maggies}).

Lensing by binary galaxies has been studied by \citet{shin08}.  From
these authors' work, we find that three image morphologies
qualitatively corresponding to im1, im2 and im3 can be obtained (see
lower panel of their Figure 10). However, five image morphologies
qualitatively corresponding to im1, im2, im3 and im4, together with a
highly demagnified fifth or central image are also possible (see
middle panel of their Figure 11).  Thus, in order to make progress, it
will be necessary to measure the redshifts of L1 and other galaxies in
the field, as well as to resolve the ambiguous status of im4.  Without
a satisfactory lensing model, any luminosity-dependent quantity
derived for the source will be uncertain by an amount corresponding to
the unknown magnification factor.

\section{Source Galaxy}
\label{sect:source}

Our X-shooter observations show images im1, im2 and im3 of the CSWA\,5
system to be gravitationally lensed images of a luminous star-forming
galaxy at $z = 1.0686$.  This is illustrated in
Figure~\ref{fig:3OIIlines}, where the individual profiles of two of
the best observed emission lines, the [O\,\textsc{ii}]\,$\lambda
\lambda 3728, 3730$ doublet and the [O\,\textsc{iii}]\,$\lambda 5008$
line are shown separately for images im1, im2 and im3.  The three sets
of profiles are indistinguishable within the noise. Furthermore, when
measured separately in each image, the redshifts of the 16 strongest
emission lines agree to within $\Delta z = 5 \times 10^{-5}$.  On the
basis of this evidence, we conclude that im1, im2 and im3 are
gravitationally lensed images of the same galaxy.

The galaxy exhibits a rich absorption and emission line spectrum which
we now describe in detail.  Unless otherwise specified, the
measurements reported below refer to the \emph{sum} of the three
images, so as to maximise the S/N ratio.

\subsection{Absorption Lines}

\begin{table}
\begin{minipage}{90mm}
     \caption{Interstellar absorption lines in the X-shooter spectrum of CSWA\,5.}
  \begin{tabular}{lllllll}
\hline
$\lambda_{\mathrm{obs}}^{\rm a}$ (\AA) & Ion  & 
$\lambda_{\mathrm{lab}}^{\rm b}$ (\AA) &
$z_{\mathrm{abs}}^{\rm a}$ & $W_0$ (\AA)  & $\log (N/{\rm cm}^{-2}$)\\
\hline
\hline
3740.44 & \ion{Si}{ii}  & 1808.013 & 1.06881 & 0.47$\pm$0.20 & 14.7\\  
3835.81 & \ion{Al}{iii} & 1854.520 & 1.06836 & 0.83$\pm$0.27 & $>$12.9\\
3853.27 & \ion{Al}{iii} & 1862.790 & 1.06855 & 0.61$\pm$0.20 & $>$13.2\\
4848.81 & \ion{Fe}{ii}  & 2344.214 & 1.06842 & 1.54$\pm$0.18 & $>$13.8\\  
4912.11 & \ion{Fe}{ii}  & 2374.461 & 1.06873 & 1.07$\pm$0.11 & 14.0\\
4928.75 & \ion{Fe}{ii}  & 2382.765 & 1.06850 & 1.14$\pm$0.09 & $>$13.0\\
5349.78 & \ion{Fe}{ii}  & 2586.650 & 1.06823 & 0.94$\pm$0.15 & $>$13.7\\
5378.10 & \ion{Fe}{ii}  & 2600.173 & 1.06836 & 1.32$\pm$0.16 & $>$13.3\\
5783.56 & \ion{Mg}{ii}  & 2796.351 & 1.06825 & 2.87$\pm$0.6 & $>$13.4\\
5798.96 & \ion{Mg}{ii}  & 2803.531 & 1.06845 & 3.06$\pm$0.7 & $>$13.7\\
5901.91 & \ion{Mg}{i}   & 2852.964 & 1.06869 & 1.23$\pm$0.6 & $>$12.4\\
\hline
\end{tabular}
$^{\rm a}$ Vacuum heliocentric. \\  
$^{\rm b}$ Vacuum rest wavelengths. \\  
\label{tab:abslines}
\end{minipage}
\end{table}

The UV-B and VIS-R spectra of the source show many strong absorption
lines from neutral and singly ionised species formed in the
interstellar medium of the lensed galaxy.  They are listed in
Table~\ref{tab:abslines}, and a few examples are reproduced in
Figure~\ref{fig:abslines}.  They define a mean absorption redshift
$z_{\rm abs} = 1.0685\pm 0.0002$, where the error quoted is the
standard deviation from the mean.  The absorption lines detected are
wide, spanning a velocity range of up to $\sim 400$\,km~s$^{-1}$.

With the relatively short exposure time devoted to CSWA\,5 during the
X-shooter commissioning, the signal-to-noise ratio of the spectra is
low in the continuum, S/N\,$\simeq 5$. Thus, only the strongest
interstellar absorption lines could be detected with confidence.  Such
lines are saturated and unsuitable for column density
determination. Under these circumstances, application of the apparent
optical depth method of \citet{savage91} yields lower limits to the
column densities $N$.  Out of the transitions listed in
Table~\ref{tab:abslines}, only two, Si\,\textsc{ii}\,$\lambda 1898$
and Fe\,\textsc{ii}\,$\lambda 2374$, may be sufficiently weak to
provide estimates of $N$(Si\,\textsc{ii}) and $N$(Fe\,\textsc{ii})
respectively. Taken at face value, the column densities of these two
ions may indicate a depletion of Fe by a factor of $\sim 5$, since Si
and Fe have comparable relative abundances in the Sun
\citep{asplund09}, whereas $N$(Fe\,\textsc{ii})\,$\simeq 1/5
N$(Si\,\textsc{ii}) in CSWA\,5.

%---------------

\begin{table}
 \begin{minipage}{80mm}
     \caption{Nebular emission lines in the X-shooter spectrum of CSWA\,5.}
  \begin{tabular}{lllll}
\hline
 $\lambda_{\mathrm{obs}}^{\rm a}$ (\AA) &  Ion & $\lambda_{\mathrm{lab}}^{\rm b}$ (\AA) & $z_{\mathrm{em}}^{\rm a}$  & Flux$^{\rm c}$ \\
\hline
\hline
3944.30~~~~~& \ion{C}{iii}]~~~~~~~~       & 1906.68~~~~~ & 1.06867~~~~~ & 0.82$\pm$0.20~~\\ 
3947.92 & \ion{[C}{iii}]       & 1908.73 & 1.06835 & 0.70$\pm$0.12\\
7710.59 & \ion{[O}{ii]}        & 3728.38 & 1.06808 & 11.24$\pm$0.06\\
7716.01 & \ion{[O}{ii]}        & 3730.29 & 1.06848 & 11.72$\pm$0.03\\
7802.38 & \ion{H}{i} (H11)     & 3769.56 & 1.06984 & 0.33$\pm$0.06\\
7858.64 & \ion{[S}{iii]}       & 3796.72 & 1.06985 & 0.25$\pm$0.04\\
7858.64 & \ion{H}{i} (H10)     & 3796.84 &  \ldots       & blend\\
7936.00 & \ion{H}{i} (H9)      & 3836.47 & 1.06857 & 0.82$\pm$0.14\\
8005.76 & \ion{[Ne}{iii}]      & 3869.86 & 1.06875 & 2.93$\pm$0.11\\
8047.27 & \ion{H}{i} (H8)      & 3890.15 & 1.06863 & 1.56$\pm$0.07\\
8047.27 & \ion{He}{i}          & 3889.75 & 1.06880 &  blend\\
8209.95 & \ion{[Ne}{iii}]      & 3968.59 & 1.06873 & 0.61$\pm$0.17\\
8215.31 & \ion{H}{i} (H7)      & 3971.20 & 1.06872 & 1.35$\pm$0.08\\
8487.74 & \ion{H}{i} (\hdelta) & 4102.90 & 1.06872 & 2.71$\pm$0.09\\
8981.69 & \ion{H}{i} (\hgamma) & 4341.69 & 1.06871 & 5.57$\pm$0.06\\
9252.58 & \ion{He}{i}          & 4472.73 & 1.06867 & 0.46$\pm$0.17\\
10059.59& \ion{H}{i} (\hbeta)  & 4862.69 & 1.06873 & 11.34$\pm$0.12\\
10261.48& \ion{[O}{iii}]       & 4960.30 & 1.06872 & 16.84$\pm$0.27\\
10369.68& \ion{[O}{iii}]       & 5008.24 & 1.06873 & 49.53$\pm$0.45\\
12159.93& \ion{He}{i}          & 5877.25 & 1.06898 & 1.20$\pm$0.15\\
13037.30& \ion{O}{i}           & 6302.05 & 1.06874 & 1.77$\pm$0.37\\
13580.46&\ion{H}{i} (\halpha)  & 6564.66 & 1.06872 & 24.12$\pm$0.50\\
13623.41&\ion{[N}{ii}]         & 6585.27 & 1.06877 & 4.33$\pm$1.13\\
14618.95& \ion{He}{i}          & 7067.14 & 1.06858 & 0.12$\pm$0.05\\
14765.79&\ion{[Ar}{iii}]       & 7137.77 & 1.06868 & 0.17$\pm$0.05\\
19722.90& \ion{[S}{iii}]       & 9533.71 & 1.06875 & 2.56$\pm$0.07\\
19797.52& \ion{H}{i} (Pa$\delta$) & 10052.2 & 1.06947 & 2.93$\pm$0.20\\
22412.11& \ion{He}{i}          & 10833.3  & 1.06882 & 1.55$\pm$0.11\\\hline
\end{tabular}
$^{\rm a}$ Vacuum heliocentric.  The redshift errors are typically
    $2\times10^{-5}$, although it the worst cases---where a feature
    is contaminated by sky residuals or telluric absorption, the redshift
    error can be up to ten times higher.\\
    $^{\rm b}$ Vacuum rest wavelengths. \\ 
$^{\rm c}$ Integrated line fluxes in units of
  $10^{-16}$ \ecs.  The errors quoted reflect only the random
  uncertainties from the counting statistics and do not include
  systematic sources of error resulting from correction for
  atmospheric absorption, subtraction of sky emission lines, and
  absolute flux calibration. For some emission lines (e.g. H$\alpha$),
  the systematic errors can be much larger than the random
  errors. \\
  \label{tab:em_lines}
\end{minipage}
\end{table}

%---------------
\subsection{Emission Lines}
\label{sec:em_lines}

\begin{figure*}
\begin{center}
{\hspace{0.15cm}\includegraphics[width=14.5cm]{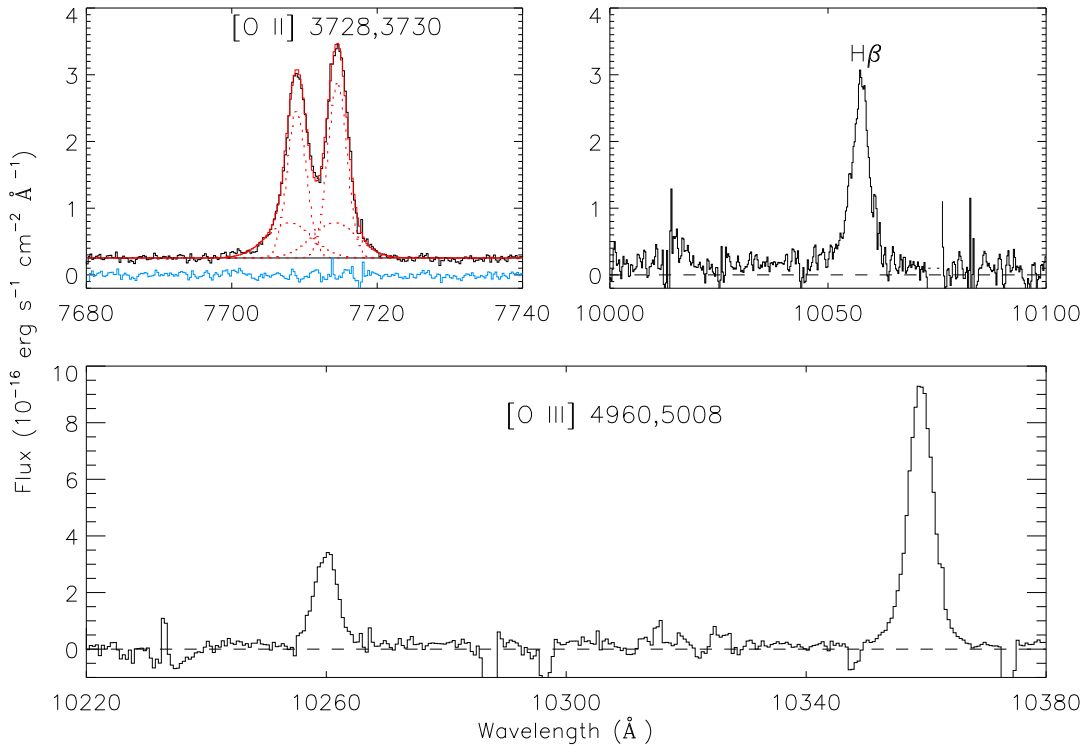}}
\end{center}
\vspace{-0.25cm}
\caption{Examples of nebular emission lines in the X-shooter spectrum
  of CSWA\,5.  The top left-hand panel illustrates the profile
  decomposition into two components for the [O\,\textsc{ii}]\,$\lambda
  \lambda 3728, 3730$ doublet. Black histogram: observed profile of
  the [O\,\textsc{ii}] emission lines. Red dotted lines: individual
  Gaussian components; each member of the doublet is fitted with two
  emission components with the parameters given in
  Section~\ref{sec:em_lines}.  Red continuous line: total fitted profile to
  the doublet.  The blue histogram near zero intensity shows the
  difference between the observed and fitted profiles.  }
\label{fig:em_lines}
\end{figure*}

Thanks to its wide wavelength coverage, the X-shooter spectrum of
CSWA\,5 includes a multitude of nebular emission lines, from
He\,\textsc{i}\,$\lambda 10833$ to the
C\,\textsc{iii}]\,$\lambda \lambda 1907, 1909$ doublet; a selection is
  reproduced in Figure~\ref{fig:em_lines}.  As can be seen from
  Table~\ref{tab:em_lines}, we detect 11 members of the Balmer series,
  as well as Pa$\delta$, and several transitions of He, C, N O, Ne, S
  and Ar in a variety of ionisation stages.  The emission lines define
  a mean redshift $z_{\rm em} = 1.06856 \pm 0.00004$ ($1 \sigma$),
  weighted by the flux in each line.

The strongest emission lines are best fit with a combination of two
Gaussian functions; a narrow component centred at $z_1=1.06859$ with a
velocity dispersion $\sigma_1 = 50 \pm 4$\,km~s$^{-1}$ (after
correction for the instrumental resolution) and a second, broader
component centred at the lower redshift $z_2=1.06843$ with $\sigma_2 =
120 \pm 10$\, km~s$^{-1}$.  The uncertainties reflect the spread of
values from acceptable fits to the line profiles. The redshift
difference corresponds to a velocity separation $\Delta v =
23$\,km~s$^{-1}$.  The top left-hand panel in
Figure~\ref{fig:em_lines} illustrates the profile decomposition for
the [O\,\textsc{ii}]\,$\lambda\lambda 3728, 3730$ doublet.
The flux ratio between the narrow and broad component varies between
1.5:1 and 2:1 among different emission lines.  The two emission
components are not easily distinguishable in the weaker, noisier
lines. Thus, in Table~\ref{tab:em_lines} we quote the integrated
fluxes across the emission lines, which are independent of profile
decomposition.

%=====================================================================
%=====================================================================

\section{Physical Properties}
\label{sect:phys_con}

The emission line fluxes collected in Table~\ref{tab:em_lines} and
their ratios allow us to deduce a number of physical parameters
characterising the H\,\textsc{ii} regions of the lensed galaxy in
CSWA\,5.  In the following, we make the assumption that the gas is
ionised by OB stars, with negligible contributions from shocks and an
active galactic nucleus (AGN; see Section~\ref{sec:bpt}). 

Physical quantities that are deduced from the ratios of emission
lines---such as reddening, gas temperature, density, and chemical
composition, or from the relative spectral energy distribution---such
as the age of the stellar population, are independent of the
magnification factor applicable to each of the three images covered by
our observations. This is not the case, however, for quantities
derived from line and continuum luminosities, such as the star
formation rate (SFR) and stellar mass ($M_{\ast}$).  In placing these
quantities on an absolute scale, we are hampered not only by the lack
of a lensing model (Section~\ref{sec:lens_model}), but also but the
unknown fraction of the light from image im3 captured by the X-shooter
slit (see Figure~\ref{fig:SDSS_im}).  To simplify matters, we used the
spectrum of image im1 alone, rather than the sum of im1, im2 and im3,
in our estimates of the star formation rate (Section~\ref{sect:sfr})
and stellar mass (Section~\ref{sec:stellar_mass}).  We chose im1
because it is less affected by blending than the other images (see
Figure~\ref{fig:SDSS_im}), and we include explicitly the unknown
magnification factor $1/f_{\rm lens}$ in our estimates of SFR and
$M_{\ast}$ below.  However, we suspect that $f_{\rm lens}$ may be of
order unity (within a factor of 2--3) \textit{for a single image}, as
found by Pettini et al. (2010) for CSWA\,20, which has a similar
configuration to CSWA\,5.

\subsection{Reddening}
\label{sec:red}

The observed ratios of Balmer line intensities can be compared with
theoretical expectations from recombination theory to deduce the
amount of dust reddening.  As is common practice in the analysis of
H\,\textsc{ii} regions, we assumed Case B recombination, an electron
temperature $T_{\rm e} = 10^4$\,K, and electron densities in the range
$n(e) = 10^2$--$10^4$\,cm$^{-3}$ \citep{osterbrock89}.  We did not
include H$\alpha$ in the analysis because at $z = 1.06856$ it is
redshifted to $\lambda_{\rm obs} = 13\,579$\,\AA, between the infrared
$J$ and $H$ bands, where atmospheric absorption is very severe and
difficult to correct for accurately (which presumably explains why the
observed ratio H$\alpha$/H$\beta$ in Table~\ref{tab:em_lines} is $\sim
25$\% \emph{lower} than the value H$\alpha$/H$\beta = 2.86$ expected
for Case B recombination).  From the observed \hbeta/\hgamma\ ratio,
we deduce $E(B-V)=0.10 \pm 0.03$, while the \hbeta/\hdelta\ ratio
gives $E(B-V) = -0.09 \pm 0.05$. The higher order Balmer lines have
larger relative line flux uncertainties, so the reddening derived from
higher order ratios is less certain. The weighted average is $E(B-V) =
0.03 \pm 0.02$.  Thus, the H\,\textsc{ii} regions of CSWA\,5 (or at
least the unobscured portions we see) suffer little dust reddening.
Since the extinction is low \citep[$A_{\rm V} \simeq 0.1$\,mag, taking
  the usual ratio of total to selective extinction $R_{V} \equiv
  A_{V}/E(B-V) \simeq 4$,][]{Calzetti00} and consistent with zero
within the uncertainties, we did not include it in the analysis
described in the following subsections.

\subsection{Star Formation Rate}
\label{sect:sfr}

We can obtain estimates of the star formation rate from the
luminosities in the H$\alpha$ emission line and the near-UV continuum,
using the calibrations of these measures by \citet{kennicutt98}.
Using the H$\beta$ flux measured for im1, $F({\rm H}\beta) = 3.1
\times 10^{-16}$\,\ecs, and the unreddened Case B recombination ratio
$F({\rm H}\alpha)/F({\rm H}\beta) = 2.86$, we deduce $F({\rm H}\alpha)
= 8.9 \times 10^{-16}$\,\ecs\ and a luminosity $L({\rm H}\alpha) = 5.1
\times 10^{42}$\,erg~s$^{-1}$.  This in turn implies:
\begin{equation}
\mathrm {SFR} = 7.9 \times 10^{-42} \, L(\halpha) \times \frac{1}{1.8}
\times \frac{1}{f_{\rm lens}} \simeq 23 ({\rm M}_{\odot}~{\rm
  yr}^{-1}),
\label{eq:sfr1}
\end{equation}
using the conversion between $L$(H$\alpha$) and SFR proposed by
\citet{kennicutt98}. The two corrections factors account for: (i) the
flattening of the initial mass function (IMF) below $1\,{\rm M}_\odot$
\citep{chabrier03} relative to the single power law of the
\citet{salpeter55} IMF assumed by \citet{kennicutt98}, and (ii) the
unknown lensing magnification of image im1, $f_{\rm lens}$, 
as discussed above.

Turning to the UV continuum, we measure $F = 6.1 \times
10^{-18}$\,\ecsa\ from our X-shooter spectrum of im1 at observed
wavelengths near 5800\,\AA.  The corresponding rest-frame luminosity
near 2800\,\AA, $L_\nu(2800) = 1.9 \times
10^{29}$\,erg~s$^{-1}$~Hz$^{-1}$, implies:
\begin{equation}
\mathrm {SFR} = 1.4 \times 10^{-28} \, L_{\nu}(2800) \times
\frac{1}{1.8} \times \frac{1}{f_{\rm lens}} \simeq 15 ({\rm
  M}_{\odot}~{\rm yr}^{-1}),
\label{eq:sfr2}
\end{equation}
with the same correction factors as above.  The two values of SFR are
in reasonable agreement, given the systematic uncertainties affecting
the two estimators, which sample different portions of the IMF
\citep[see][]{meurer09}.

\subsection{Electron Temperature and Density}
\label{sect:eldens}

At the S/N ratio of the present data, the auroral lines most commonly
used for the determination of the electron temperature,
[\ion{O}{iii}]\,$\lambda 4363$, [\ion{N}{ii}]\,$\lambda 5755$, and
[\ion{S}{iii}]\,$\lambda 6312$ are below the detection limit.  The $ 3
\sigma$ upper limit $F(4363) \leq 0.3 \times 10^{-16}$\ecs\ translates
to an upper limit on the temperature $T_{\rm e} \leq 9900$\,K
\citep{aller84}. 
Turning to density sensitive line
ratios, we resolve both the [\ion{O}{ii}]\,$\lambda \lambda 3728,
3730$ and the C\,\textsc{iii}]\,$\lambda \lambda 1907, 1909$ doublets,
  while [\ion{S}{ii}]\,$\lambda \lambda 6716, 6731$ falls between the
  $J$ and the $H$ bands, where the atmospheric transmission is a few
  percent only.  We used the \textsc{nebular} package in
  \textsc{iraf}, incorporating the calculations by \citet{shaw95}, to
  deduce $n(e) = 315 \pm 15$\,cm$^{-3}$ from the observed 
  $F(3726)/F(3728)$ ratio, assuming $T_{\rm e} = 10\,000 \pm 1000$\,K.
  Although less tightly constrained, the observed $F(1907)/F(1909)$
  ratio is consistent with the electron density deduced from the
  higher S/N ratio [O\,\textsc{ii}] doublet.

\begin{figure*}
\begin{center}
\includegraphics[width=16.5cm]{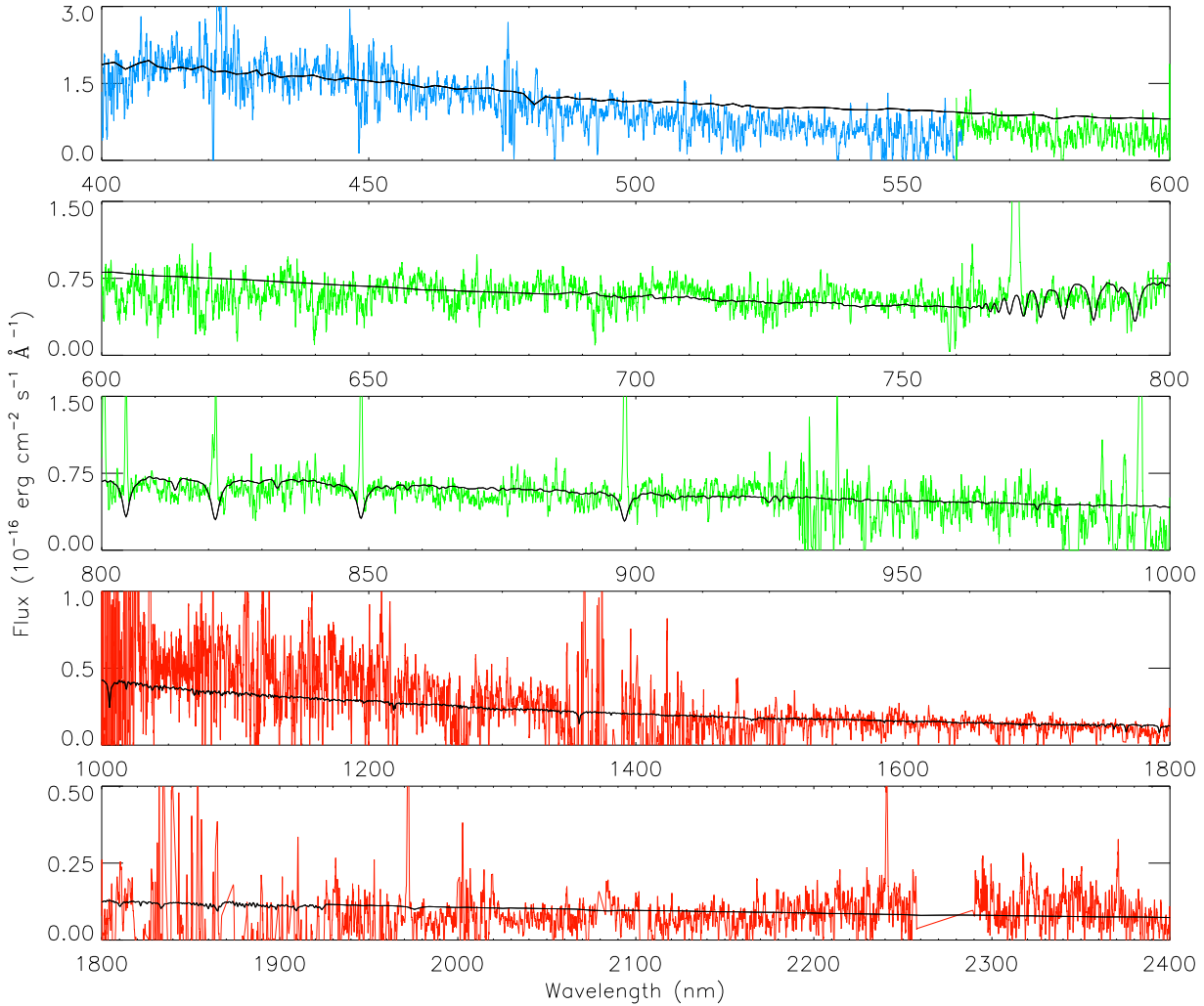}
\end{center}
\vspace{-0.25cm}
\caption{X-shooter spectrum of image im1 in CSWA\,5, colour-coded to
  show the spectral ranges covered by the three arms of the
  spectrograph, as follows: blue is UV-B, green is VIS-R, and red is
  NIR.  The spectrum has been smoothed with a boxcar filter of width
  2.5\,\AA.  Overplotted in black is the best-fitting template
  spectrum computed with a \citet{bruzual03} model for a 40\,Myr-old
  continuous star formation episode and assembled stellar mass $M_\ast
  = 2.8 \times 10^{9} {\rm M}_\odot$.  }
\label{fig:BC_fit}
\end{figure*}

\subsection{Chemical Abundances}
\label{sec:abundances}

With many nebular lines detected, we can estimate the abundances of
the C, N, O group using a variety of strong emission line diagnostics.
Results are collected in Table~\ref{tab:abund}.

\vspace*{0.1cm} {\flushleft \textit{Oxygen.} \hspace{0.0025cm} }
Oxygen is the element most commonly used to characterise the overall
degree of metal enrichment of H\,\textsc{ii} regions and we thus
consider it first.  Among the many calibrations of the oxygen
abundance based on the ratios of strong emission lines
\citep[e.g.][]{kewley08}, the $R23$ method first proposed by
\citet{pagel79} is the one still most commonly applied, at least to
low redshift galaxies.  We used the calibration of the index $R23
\equiv [F(3728) + F(3730) + F(4960) + F(5008)]/F({\rm H}\beta)$ by
\citet{kobulnicky04}, appropriate to the lower branch of the
double-valued $R23$ vs. O/H relation, to deduce $12 + \log{\rm
  (O/H)}_{R23} = 8.28$, or approximately 40\% of the oxygen
abundance in the Sun \citep[][]{asplund09} and in the Orion nebula
\citep[][]{esteban04}.  The random error in the value of $\log{\rm
  (O/H)}$ is small and negligible compared with the 0.2--0.3\,dex
systematic uncertainty in the calibration of the $R23$ index
\citep{kewley08}. The lower branch solution is favoured by the low
value of the N abundance deduced below.

The calibrations of the oxygen abundance based on the ratios of
nitrogen, oxygen and Balmer lines proposed by \citet{pettini04} have
the advantages, compared to $R23$, of being single-valued and
relatively robust to errors in flux calibration and dust extinction.
However, in the case of CSWA\,5, their use is possibly compromised by
the awkward redshifted wavelengths of H$\alpha$ and the
[N\,\textsc{ii}]\,$\lambda \lambda 6550, 6585$ doublet which fall
between the $J$ and $H$ bands, as already explained.  We deduce $12 +
\log{\rm (O/H)}_{\rm N2} = 8.47 \pm 0.07$ from consideration of the
ratio $N2 \equiv \log [F(6585)/F(H\alpha)]$, and $12 + \log{\rm
  (O/H)}_{\rm O3N2} = 8.29 \pm 0.04$ from the ratio $O3N2 \equiv \log
[F(5008)/F(H\beta)] - \log [F(6585)/F(H\alpha)]$, using in both cases
the calibrations of these indices deduced by \citet{pettini04}. Again,
the random errors quoted are small compared to the 0.2--0.3 dex
systematic uncertainties of the calibrations.  We conclude that the
oxygen abundance in the lensed galaxy in CSWA\,5 is $12+ \log {\rm
  (O/H)} = 8.3$--8.5, or (O/H)$_{\rm CSWA\,5} \simeq 0.4$--$0.65 \,
{\rm (O/H)}_\odot$.

\begin{table}
 \centering
 \begin{minipage}{62mm}
     \caption{C, N, O Abundances in CSWA\,5.}
 \begin{tabular}{ccc}
\hline
$12 + \log{\rm (O/H)}$  & ~~~~~$\log {\rm (N/O)}$   & ~~~$\log {\rm (C/O)}$ \\
\hline
\hline
8.3--8.5 &  $-1.0 \pm 0.1$ & $-0.6 \pm 0.2$ \\
\hline
\end{tabular}
\label{tab:abund}
\end{minipage}
\end{table}

%=========
\vspace*{0.1cm} {\flushleft \textit{Nitrogen.} \hspace{0.0025cm} } In
the absence of temperature-sensitive auroral lines, we make use of the
method developed by \citet{thurston96} to deduce the relative
abundances of N and O.  These authors used photoionization models to
deduce a relationship between the [N\,\textsc{ii}] temperature and the
$R23$ index; in our case, $t_{[N\,\textsc{ii}]} = 11\,000 \pm
1000$\,K.  Once the [N\,\textsc{ii}] temperature is known, the N/O
ratio can be deduced directly from the ratio of the
[N\,\textsc{ii}]\,$\lambda \lambda 6550, 6585$ and
[O\,\textsc{ii}]\,$\lambda \lambda 3728, 3730$ doublets
\citep{pagel92}, under the assumption that N$^+$/O$^+ \simeq$\,N/O
which \citet{thurston96} argue introduces only a small error.
Following this procedure, we find $\log {\rm (N/O)}_{\rm CSWA\,5} =
-0.95 \pm 0.10$ or $\sim 80$\% of the solar value $\log {\rm
  (N/O)}_\odot = -0.86$ \citep{asplund09}.  A slightly subsolar N/O
ratio when the abundance of oxygen is $\sim 0.5$ solar is in line with
the large body of such measurements now available in galaxies at a
range of redshifts \citep[e.g.][]{izotov06, pettini08b}.

%=========

\vspace*{0.1cm} {\flushleft \textit{Carbon.} \hspace{0.0025cm} } In
the low density limit, the ratio C$^{+2}$/O$^{+2}$ can be deduced from
the measured fluxes of the C\,{\textsc{iii}]\,$\lambda \lambda 1907, 1909$ and
[O\,\textsc{iii}]\,$\lambda \lambda 4960, 5008$ doublets and a knowledge of
the relevant collision strengths. Using the line emissivities given by
\citet{aller84}, we deduce $\log ({\rm C}^{+2}/{\rm O}^{+2}) = -0.6
\pm 0.2$, for $T_{\rm e} = 10\,000 \pm 1000$\,K.  The ionization
correction factor to deduce C/O from C$^{+2}$/O$^{+2}$ is expected to
be small \citep{garnett95}, and therefore we find $({\rm C}/{\rm
  O})_{\rm CSWA\,5} \simeq 0.3$--$0.7 ({\rm C}/{\rm O})_{\odot}$,
where $\log ({\rm C}/{\rm O})_\odot = -0.26$ \citep{asplund09}.  As
for nitrogen, the sub-solar C/O ratio in CSWA\,5 conforms to the
established behaviour of C/O vs. O/H in metal-poor H\,\textsc{ii}
regions \citep{garnett95, garnett99}, Galactic stars \citep{akerman04,
  fabbian09}, and Lyman break galaxies \citep{Shapley03,erb10}.

\subsection{Age and Assembled Stellar Mass}
\label{sec:stellar_mass}

Finally, we can obtain estimates of the age and stellar mass of the
lensed galaxy by comparing its observed spectral energy distribution
(SED), from the rest frame UV to the near-IR (from $\sim
2000$\,\AA\ to $\sim 1.2 \mu$m where our X-shooter spectrum has the
highest S/N ratio), to those of synthetic spectra computed with simple
stellar population models \citep[e.g.][]{erb06a}.  For this purpose,
we used the population synthesis code of \citet{bruzual03} with
\citet{chabrier03} IMF, metallicity $Z = 0.4 Z_\odot$
(Section~\ref{sec:abundances}), and no internal reddening
(Section~\ref{sec:red}).  We generated two families of models with
these parameters, for the two limiting cases of continuous star
formation and an instantaneous burst, in each case varying the age
from 1 to 70\,Myr.  The best fitting models and the corresponding
values of age and assembled stellar mass were determined by minimizing
the value of $\chi^2$, given by:
\begin{equation}
\chi^2= 
\sum_\lambda 
\frac{(f_{\mathrm{obs},\lambda} - b \times f_{\mathrm{model}, \lambda})^2}
{\sigma_{\mathrm{obs}, \lambda}^2} 
\label{eq:chi2}
\end{equation}
where $f_{\mathrm{obs}, \lambda}$ and $\sigma_{\mathrm{obs}, \lambda}$
are the observed flux at wavelength $\lambda$ and its error
respectively, and $f_{\mathrm{model}, \lambda}$ is the flux of the
\citet{bruzual03} model spectrum at the same wavelength.  The
normalization factor $b$ gives the stellar mass of the galaxy.

In fitting the model spectra to the X-shooter spectrum of image im1,
we excluded regions affected by strong telluric absorption or by
prominent residuals in the subtraction of sky emission lines. We found
that the best fitting models (see Figure~\ref{fig:BC_fit}) are those
with ages of $40 \pm 10$\,Myr and stellar masses $M_\ast = (2.8 \pm
1.0) \times 10^9 \times 1/f_{\rm lens} \, {\rm M}_\odot$, where
$f_{\rm lens}$ is the unknown magnification factor of im1, and the
error includes the 20\% uncertainty in the absolute flux calibration
(Section~\ref{sect:data}).

These values appear to be robust to the choice of star formation mode,
with the single burst and continuous star formation models converging
to similar solutions.  We also investigated the possibility that the
stellar continuum suffers a greater extinction than the H\,\textsc{ii}
emission lines, since the former could in principle sample a different
(presumably older) stellar population than the latter.  However,
without recourse to the rest frame far-UV spectral range (which is
inaccessible from the ground at $z = 1.0686$), we come up against the
well-known age-extinction degeneracy. For example, models in which the
stellar continuum is reddened with a colour excess $E(B-V) = 0.3$ also
provide satisfactory fits to the X-shooter spectrum of CSWA\,5, albeit
with younger preferred ages of $25 \pm 7$\,Myr and lower stellar
masses of $(2.3 \pm 0.7) \times 10^9 \times 1/f_{\rm lens} \, {\rm
  M}_\odot$.

We can obtain an independent estimate of the age of the starburst from
consideration of the equivalent width of the H$\beta$ emission line,
$W_0({\rm H}\beta) \simeq 100$\,\AA, using the values of $W_0({\rm
  H}\beta)$ as a function of time calculated with the
\textsc{Starburst99} spectral synthesis code \citep{leitherer01}.  The
equivalent widths of the Balmer recombination lines fall rapidly
following a burst of star formation and $W_0({\rm H}\beta) <
100$\,\AA\ for all ages greater than 5\,Myr.  On the other hand, the
time dependence is less steep for continuous star formation and ages
of $\sim 25$\,Myr are indicated, in better agreement the age inferred
from the analysis of the UV to near-IR SED.

\section{Discussion}
\label{sec:discuss}

Summarizing our findings, we have established that CSWA\,5 is an
actively star-forming galaxy at $z = 1.0686$ lensed by an apparent
foreground group of massive red galaxies, at least one of which is at
$z = 0.3877$.  The lensed source is forming stars at a rate ${\rm SFR}
\simeq 20\, {\rm M}_\odot$~yr$^{-1}$, suffers negligible reddening,
has an oxygen abundance of approximately half-solar, and sub-solar N/O
and C/O ratios (as expected for low metallicity galaxies).  Its blue
spectral energy distribution, from the rest-frame UV to the near-IR,
and high H$\beta$ equivalent width are indicative of a young age, only
$\sim 25$--50\,Myr, during which time the galaxy has assembled a
stellar mass of $\sim 3 \times 10^9 {\rm M}_\odot$.  The
star-formation rate and stellar mass are uncertain by an unknown
magnification factor, which may be of order unity for each of the four
gravitationally lensed images of CWSA\,5.

These physical characteristics are broadly in line with those of the
population of UV-selected galaxies at redshifts $z = 1$--3
\citep[e.g.][and references therein]{pettini07}, as we now discuss.

\subsection{Mass-Metallicity Relation}
\label{sec:M-Z}

The most relevant comparison here is with the mass-metallicity
relation determined by \citet{savaglio05} for galaxies at a mean
$\langle z \rangle \simeq 0.7$ selected from the Gemini Deep Deep
Survey (GDDS) and shown in Figure~\ref{fig:massmetal}.  It must be
borne in mind here that such comparisons are fraught with pitfalls for
the unwary, because of systematic differences between the methods used
to determine both the stellar mass and the metallicity. In this
particular case, the study by \citet{savaglio05} should be compatible
with ours, as both used a \citet{chabrier03} IMF in arriving at the
stellar masses and the $R23$ index for the oxygen abundance
\citep[although the reliability of the latter at apparently
  super-solar metallicities is questionable---see][]{pettini08a}.

An additional complication is the unknown magnification factor which
applies to our derivation of the stellar mass. If $f_{\rm lens}$ in
Section~\ref{sec:stellar_mass} is of order unity, then
Figure~\ref{fig:massmetal} shows that CSWA\,5 is somewhat metal-poor
for its stellar mass, compared to GDDS galaxies at $z \sim 0.7$.
Definite conclusions are difficult at this stage, given the scatter of
the GDDS galaxies about the mean relation in
Figure~\ref{fig:massmetal}, and the lack of objects as metal-poor as
CSWA\,5 in the sample considered by \citet{savaglio05}.  Taken at face
value, the offset of CSWA\,5 in Figure~\ref{fig:massmetal} is in the
same sense as the more general redshift evolution of the
mass-metallicity relation proposed by \citet{maiolino08}.

\begin{figure}
\begin{center}
\vspace*{-0.25cm}
{\hspace*{-1cm}\includegraphics[width=9cm]{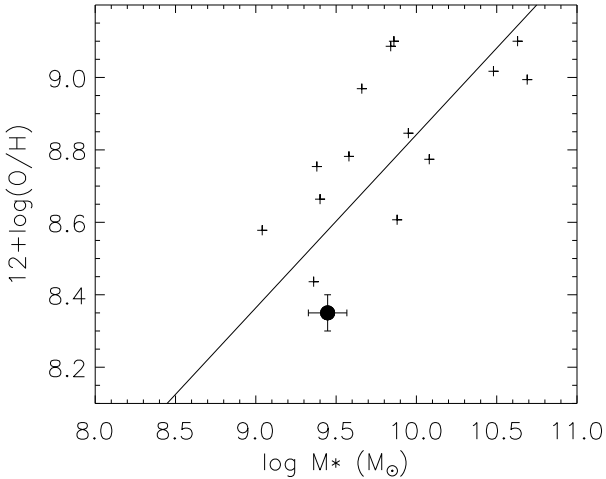}}
\end{center}
\vspace{-0.5cm}
\caption{Mass metallicity relation for galaxies at $z\sim0.7$ from the
  GDDS \citep[crosses;][]{savaglio05}. The location of CSWA\,5 is
  indicated by the large filled circle. The errors shown do not
  include the systematic uncertainties in the metallicity and mass
  determinations.  }
\label{fig:massmetal}
\end{figure}

\subsection{Specific Star Formation Rate}

Since the lens magnification factor enters into the calculation of
both SFR and $M_\ast$, the ratio of these two quantities, commonly
referred to as the specific star formation rate (SSFR), should not be
affected.  For CSWA\,5 we deduce a specific ${\rm SFR} = (8.2 \pm 2.2)
\times 10^{-9}$\,yr$^{-1}$.  The reciprocal of this value gives a
timescale for the star formation activity of $\sim 120 \pm 45$\,Myr,
which is $\sim 3$ times higher than the age deduced in
Section~\ref{sec:stellar_mass} from SED fitting.

In Figure~\ref{fig:ssfr} we compare the value of SSFR in CSWA\,5 with
those of UV-selected galaxies at $z \simeq 2$ from the work by
\citet{erb06a} and of lower redshift galaxies from the VVDS survey by
\citet{lamareille09}. Clearly, CSWA\,5 is more akin to the former than
to the latter in its relatively high SSFR and correspondingly short
star formation timescale.

\subsection{Emission line ratios}
\label{sec:bpt}

Figure~\ref{fig:bpt} shows that CSWA\,5 is well separated from the
locus of SDSS galaxies in the ionization diagnostic diagram of
\citet{baldwin81}, sometimes referred to as the BPT diagram.  While we
find no evidence for the presence of an AGN in CSWA\,5 from the width
of the emission lines, we note that such offsets now appear to be
common for the most actively star-forming galaxies \citep{erb06c,
  liu08}.  \citet{brinchmann08} pointed out that, among SDSS galaxies,
there is a correlation between excess SSFR and offset from the main
locus occupied by `normal' galaxies in the BPT diagram.  Galaxies with
higher SSFRs tend to exhibit higher values of the ratios
[O\,\textsc{iii}]/H$\beta$ and [N\,\textsc{ii}]/H$\alpha$.  CSWA\,5
apparently fits this trend.  The underlying physical explanation for
the offset is not well established, but there are indications that it
may be related to a higher ionization parameter of the H\,{\sc ii}
regions, which in turn is presumably linked to the high star formation
activity \citep[][]{liu08, brinchmann08}.

\begin{figure}
\begin{center}
\includegraphics[width=8cm]{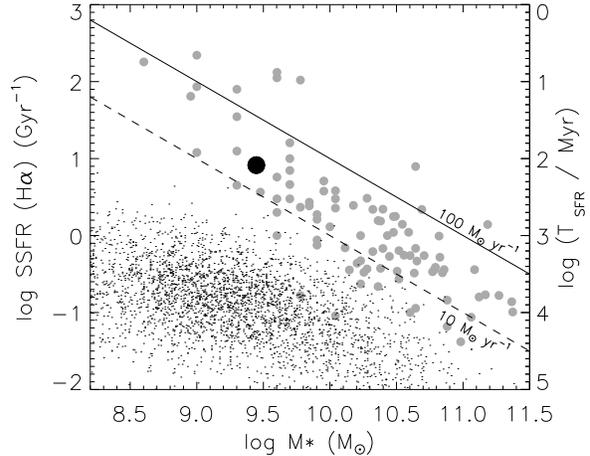}
\end{center}
\caption{Specific star-formation rate of CSWA\,5 (large black dot)
  compared with those of galaxies at $z>2$ \citep[grey
    dots;][]{erb06a}, and galaxies at $0<z<0.6$ from the VIMOS VLT
  Deep Survey (VVDS) \citep[small dots;][]{lamareille09}.  All SFRs
  were derived from the \halpha\ luminosities.  Lines of constant SFR
  are indicated by the straight solid and dashed lines.  The mass (but
  not the specific SFR) of CSWA\,5 is uncertain by an unknown factor
  $1/f_{\rm lens}$ which may be of order unity.  The scale on the
  right vertical axis shows the formation time scale (the inverse of
  the specific SFR).}
\label{fig:ssfr}
\end{figure}

%%%%%%%%%%%%%%%%%%%%%%%%%%%%%%%%%%%%%%%%%%%%%%%%%%%%%%%%%%%%%%%%%%%%%%
\section{Conclusions}
\label{sec:conclude}

It is encouraging and satisfying that through dedicated searches of
large area surveys the number of gravitationally lensed galaxies at
high redshifts continues to increase.  The work presented here has not
only identified a new such example in the CASSOWARY catalogue but,
more importantly, has demonstrated the wealth of information on the
physical conditions of these objects which can be gleaned from even
relatively short exposures with the X-shooter spectrograph, thanks to
its high efficiency and wide wavelength coverage.  The main impediment
we have found to the full characterization of CSWA\,5 is the lack of a
lensing model which can reproduce the deflections of the images.
Higher spatial resolution imaging of the field (achievable with the
\textit{Hubble Space Telescope}) and more redshift determinations of
galaxies in the field are needed for a comprehensive census of the
mass distribution along the line of sight and an accurate estimate of
the magnification of CSWA\,5.  Such information, combined with higher
S/N ratio spectra of the four lensed images (the work presented in
this paper is based on only 40 minutes of integration on target),
would lead to a full mapping of the properties of this star-forming
galaxy.

As a final note, we should like to stress that the use of
gravitational lensing to access the properties of galaxies at high
redshifts is in some way still in its infancy.  While the few cases
studied so far have been very successful in providing detailed views
of these systems which would otherwise have simply been unobtainable,
it is also the case that nearly all of them appear to be drawn from
the bright end of the galaxy luminosity function. Fainter galaxies
\textit{are} accessible with 8--10\,m telescopes equipped with modern
spectrographs, but are often overlooked in searches for strong
gravitational lenses.  By systematically extending such searches to
fainter limits with current and future (e.g. with the Large Synoptic
Survey Telescope) surveys, it will be possible to identify
intrinsically fainter galaxies, particularly at epochs when the
Universe was forming most of its stars, and thereby assemble a full
picture of the galaxy formation process in action.

\begin{figure}
\begin{center}
\includegraphics[width=7.5cm, bb=69 365 410 658, clip]{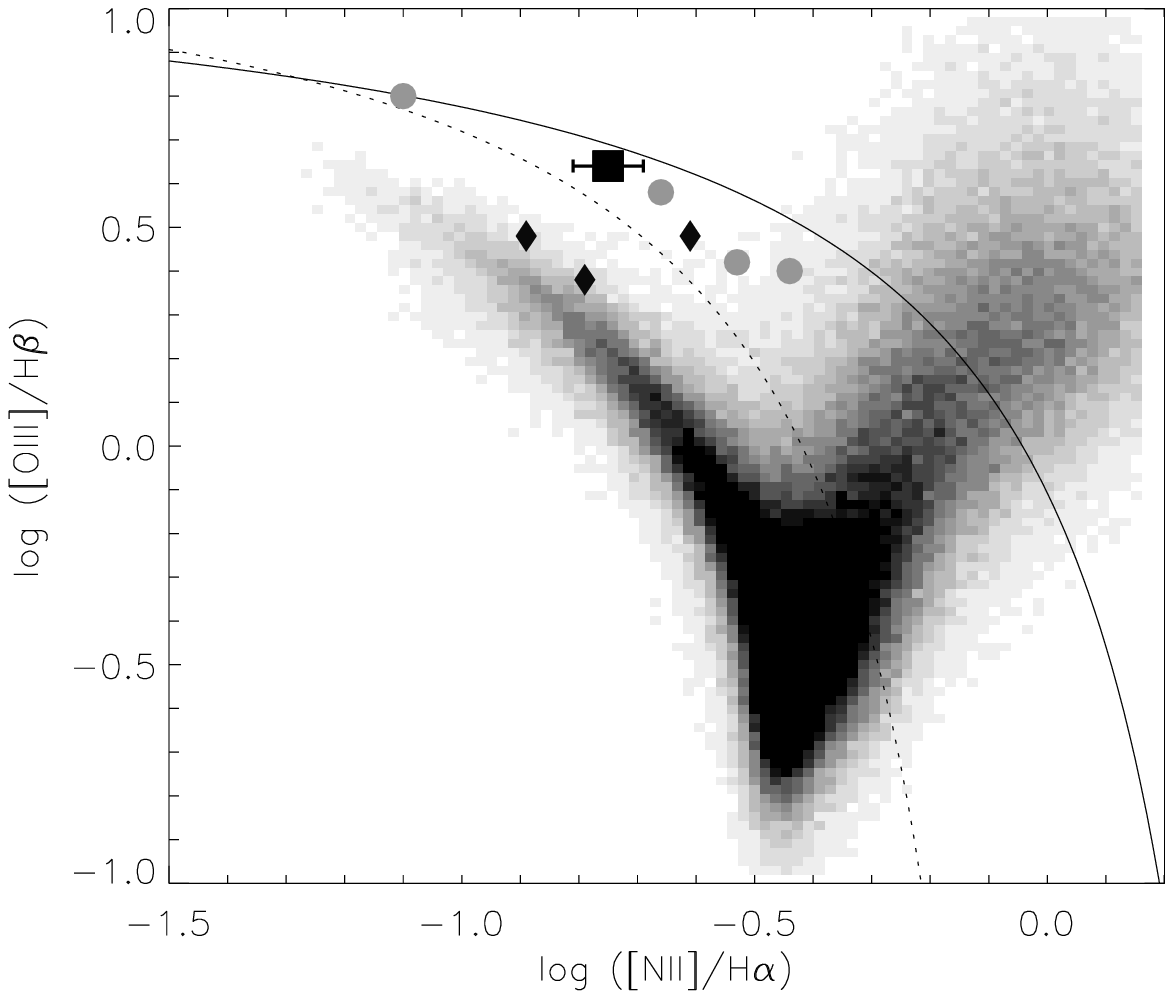}
\end{center}
\caption{Emission line flux ratios of SDSS galaxies shown in grey
  scale \citep{tremonti04}. The solid and dotted lines represent the
  theoretical \citep{kewley01} and empirical \citep{kauffmann03}
  demarcations between star forming galaxies and galaxies harbouring
  AGN.  Also overplotted are the values for CSWA\,5 (large square),
    and those measured from spectra of star-forming
    galaxies at $z \simeq 2$ \citep[][circles]{erb06c}, and
    $z \simeq 1$--1.4 \citep[][diamonds]{liu08} .} 
\label{fig:bpt}
\end{figure}

\section*{Acknowledgments}
The good quality of the spectra obtained during the commissioning runs
of the instrument was the result of the dedicated and successful
efforts by the entire X-shooter Consortium team.  More than 60
engineers, technicians, and astronomers worked over more than five
years on the project in Denmark, France, Italy, the Netherlands, and
at ESO.  S.D. would like to acknowledge, in representation of the
whole team, the co-Principal Investigators P.~Kj{\ae}rgaard Rasmussen,
F.~Hammer, R.~Pallavicini, L.~Kaper, and S.~Randich, and the Project
Managers H.~Dekker, I.~Guinouard, R.~Navarro, and F.~Zerbi.  Special
thanks go to the ESO commissioning team, in particular H.~Dekker,
J.~Lizon, R.~Castillo, M.~Downing, G.~Finger, G.~Fischer, C.~Lucuix,
P.~Di~Marcantonio, E.~Mason, A.~Modigliani, S.~Ramsay and P.~Santin.
We are grateful to the anonymous referee for useful suggestions that
improved the presentation of the work.  M.P. would like the express
his gratitude to the members of the International Centre for Radio
Astronomy Research at the University of Western Australia for their
generous hospitality during the completion of this work.

%----------------------
%\bibliographystyle{mn2e}
\bibliographystyle{apj}
\bibliography{ms_lc}
%----------------------
\end{document}